\documentclass[prl,amsmath,superscriptaddress,twocolumn,showpacs]{revtex4-1}
\usepackage{bm}
\usepackage{amssymb}
\usepackage{colordvi}
\usepackage{graphicx}
\usepackage{color}
\usepackage{hyperref}

\newcommand{\beq}{\begin{equation}}
\newcommand{\eeq}{\end{equation}}
\newcommand{\bea}{\begin{eqnarray}}
\newcommand{\eea}{\end{eqnarray}}

\newcommand{\vep}{\varepsilon}

\begin{document}

\title{Topological light-trapping on a dislocation}

\author{Fei-Fei Li}\email{These authors contributed equally.}
\affiliation{School of Electronic Science and Engineering, Nanjing University, Nanjing, 210093, China}
\author{Hai-Xiao Wang}\email{These authors contributed equally.}
\affiliation{College of Physics, Optoelectronics and Energy, \&
  Collaborative Innovation Center of Suzhou Nano Science and
  Technology, Soochow University, 1 Shizi Street, Suzhou 215006,
  China}
\author{Zhan Xiong}\email{These authors contributed equally.}
\affiliation{College of Physics, Optoelectronics and Energy, \&
  Collaborative Innovation Center of Suzhou Nano Science and
  Technology, Soochow University, 1 Shizi Street, Suzhou 215006,
  China}
\author{Qun Lou}
\affiliation{School of Electronic Science and Engineering, Nanjing University, Nanjing, 210093, China}
\author{Ping Chen}
\affiliation{School of Electronic Science and Engineering, Nanjing University, Nanjing, 210093, China}
\author{Rui-Xin Wu}
\affiliation{School of Electronic Science and Engineering, Nanjing University, Nanjing, 210093, China}
\author{Ying Poo}\email{ypoo@nju.edu.cn}
\affiliation{School of Electronic Science and Engineering, Nanjing University, Nanjing, 210093, China}
\author{Jian-Hua Jiang}\email{jianhuajiang@suda.edu.cn, joejhjiang@hotmail.com}
\affiliation{College of Physics, Optoelectronics and Energy, \&
  Collaborative Innovation Center of Suzhou Nano Science and
  Technology, Soochow University, 1 Shizi Street, Suzhou 215006,
  China}
\author{Sajeev John}
\affiliation{College of Physics, Optoelectronics and Energy, \&
  Collaborative Innovation Center of Suzhou Nano Science and
  Technology, Soochow University, 1 Shizi Street, Suzhou 215006,
  China}
\affiliation{Department of Physics, University of Toronto, 60 Saint George Street, Toronto, Ontario, M5S 1A7, Canada}

\date{\today}

\maketitle


{\bf Topology has been revealed to play a fundamental role in physics in the past decades \cite{a1,a2,a3,a33,a4,a5,a6}. Topological insulators have unconventional gapless edge states where disorder-induced back-scattering is suppressed \cite{a4,a5,a6}. In photonics, such edge states lead to unidirectional waveguides which are useful for integrated photonic chips \cite{b1,b2,b3,b4,b5,b6,b7,b8,c1,c2,c3,d1,d2,d3,d4,d5,d6,d7,d8,e1,e2,e3,f1,f2,f3}. Cavity modes, another type of fundamental components in photonic chips, however, are not protected by band topology because of their lower dimensions. Here we demonstrate that concurrent wavevector-space and real-space topology, dubbed as the ``dual-topology", can lead to light-trapping in lower-dimensions. The resultant photonic bound state emerges as a Jackiw-Rebbi soliton mode \cite{a1} localized on a dislocation in a two-dimensional (2D) photonic crystal, as predicted theoretically and discovered experimentally. Such a strongly-confined 0D localized mode, which is solely due to the topological mechanism, is found to be robust against perturbations. Our study unveils a new mechanism for topological light-trapping in lower-dimensions, which is valuable for fundamental physics and a variety of applications in photonics.}

Photonic crystals (PCs) are periodic structures of electromagnetic (EM) materials which offer versatile tailoring of photonic spectrum and wave dynamics \cite{g1,g2,g3}. Recently, photonic quantum anomalous Hall effects \cite{b1,b2,b3,b4,b5,b6,b7,b8}, photonic Floquet topological insulators \cite{c1,c2,c3}, photonic quantum spin Hall insulators \cite{d1,d2,d3,d4,d5,d6,d7,d8}, topological photonic quasicrystals \cite{e1,e2,e3} and photonic Zak phases \cite{f1,f2,f3} are realized or proposed using various PCs. Topology is revealed as a mechanism for light-trapping on the edges of PCs \cite{b1,b2,b3,b4,b5,b6,b7,b8,c1,c2,c3,d1,d2,d3,d4,d5,d6,d7,d8,e1,e2,e3,f1,f2,f3}, leading to topological surface states which are much more robust than conventional surface states \cite{g3}. However, up till now, topological light-trapping at (sub-)wavelength scale is achieved only for edges which have one less dimension than the bulk. In this Letter, we predict theoretically and observe experimentally robust light-trapping into a zero dimensional (0D) cavity mode in a 2D PC, as induced and protected by the dual-topology mechanism. The characteristic localization length $l_{loc}$ of the cavity mode is close to the wavelength in vacuum $\lambda$. Strong and robust light-trapping is ubiquitously useful in the state-of-art photonics such as miniature photonic devices, optical sensors, and cavity quantum electrodynamics. We emphasize that although the dual-topology mechanism was first proposed in electronic systems \cite{j1,j2,j3}, it is much more feasible to observe and utilize dual-topology to induce wave localization in photonics than in electronics. To date, such wave localization into $d-2$ dimensions in a $d$-dimensional system has not yet been discovered experimentally in either realms.

The proposed structure is a rectangular-lattice PC with spatial periodicities $a_x$ and $a_y$ along the $x$ and $y$ directions, respectively. The photonic band gap (PBG) has Chern number ${\cal C}=1$ and the Zak phase \cite{a33} along the Brillouin zone (BZ) boundary line XM is $\theta_{XM}=\pi$. These two special properties constitute the nontrivial topology in wavevector-space. On the other hand, a dislocation is an object with real-space topology: any close-loop of lattice translation including the dislocation has a mismatch between the starting and ending points. The vector connecting these points, the Burgers vector, serves as the real-space topological index (see Fig. 1a).

To elucidate the light-trapping mechanism through the dual-topology, we employ the ``cut-and-glue" technique \cite{j1} (see Figs. 1b and 1c) which consists of two steps: first, a chunk of PC with trivial topology is inserted into the dislocation structure which is then split into two halves. This step introduces two one-way edge channels at the opposite boundaries of the chunk, due to the wavevector-space topology. The dispersions of these two edge channels must intersect at a time-reversal invariant wavevector. The nontrivial Zak phase along the XM line ensures such an intersection to be at $K_x=\frac{\pi}{a_x}$ \cite{j1,f1}. For a finite-width chunk, the tunnel coupling between the opposite edges opens a gap at the intersection (Dirac) point. These two coupled edges can be described by the 1D massive Dirac Hamiltonian, ${\cal H}=vq_x\sigma_z+m\sigma_x$, where $v$ is the group velocity of the edge states, $\sigma_z=\pm 1$ represents the two edge channels and $q_x$ is the wavevector relative to the Dirac point. The Dirac mass $m$ ($m$ is real, see Supplemental Materials) depends on the inter-edge coupling. In the weak coupling regime, the Dirac mass is determined by the overlapping integral of the EM fields of the two edge states \cite{g3}, i.e., $m \propto \int d{\vec r} ({\vec E}_+\cdot \hat{\vep}\cdot {\vec E}_{-}^\ast + {\vec H}_+ \cdot  \hat{\mu}\cdot {\vec H}_{-}^\ast + {\rm c.c.})$ where the subscripts $\pm$ represent the edge states at the upper and lower boundaries, respectively. Due to the dislocation, the Dirac mass becomes position-dependent, since the upper and lower edges experience different numbers of lattice translations (see Fig. 1b). If the Burgers vector is ${\vec B}=(a_x, 0)$, the phase difference between the two edge states will experience a $\pi$-phase abrupt across the dislocation, since $K_x a_x=\pi$. This $\pi$-phase abrupt leads to a sign-change in the Dirac mass, forming a mass domain-wall at the dislocation. According to the Jackiw-Rebbi theory \cite{a1}, there will be a mode localized at the dislocation in the PBG. 

The second step in the cut-and-glue procedure is to glue the two halves together by reducing the width of the chunk gradually to zero. As the width of the chunk reduces, the inter-edge coupling becomes stronger and stronger. The magnitude of the Dirac mass increases and the edge states are gradually moved into the bulk bands. However, because the Dirac mass domain-wall persists due to the real- and wavevector-space topology, the Jackiw-Rebbi soliton mode always exists in the PBG, even when the chunk of trivial PC is removed (see Supplemental Materials for the evolution of the Jackiw-Rebbi soliton mode and the edge states during the cut-and-glue processes using EM wave simulation).

The above scenario of light-trapping by the dual-topology mechanism can be mathematically summarized as \cite{j1,j2,j3}, 
\beq
N_{loc}=\frac{1}{\pi}{\vec B}\cdot {\vec Q} : {\rm mod}~2,
\eeq
where ${\vec Q}=(\frac{\theta_{XM}}{a_x},\frac{\theta_{YM}}{a_y})$ with $\theta_{XM}$ and $\theta_{YM}$ being the Zak phases along the XM and YM lines, respectively, and $N_{loc}=0$ or 1 is the number of localized photonic modes. The existence of the mass domain-wall and the soliton mode has a $Z_2$ nature, reflecting whether the Dirac mass switches sign or not. We emphasize that the Chern number, though not appeared in the above equation, provides the ground for the edge states and the Jackiw-Rebbi soliton \cite{dis1,dis2}. This crucial requirement differs our mechanism from those in Refs.~\cite{pdis1,pdis2}, beside the strong wave localization observed in this work.

To realize light-trapping due to the dual-topology mechanism, we design a rectangular-lattice PC with two yttrium iron garnet (YIG) cylindrical pillars in each unit cell. With the metallic cladding above and below, the bulk photonic bands of interest here are the 2D transverse-magnetic (TM) harmonic modes. Magnetized by a magnetic field of 900~Oe along the $z$ direction, a PBG (denoted as ``PBG II") between the third and fourth bands with Chern number ${\cal C}=1$ is developed for $a_y=2a_x=24$~mm, $R=2$~mm and $d=17$~mm. This topological PBG is realized by the gyromagnetic effect which gaps out the two Dirac points (indicated by the red arrow) on the YM line (see Fig.~2a). The dispersions along the BZ boundary lines YM and XM are shown in Fig. 2b together with the Zak phases along these lines. The Zak phases are calculated using the Wilson-loop approach (see Supplemental Materials). The EM field profiles at high symmetry points indicate connection and consistency between our ``diatomic" photonic unit-cell and the Su-Schrieffer-Heeger model \cite{a2}: the parity inversion between the X and M points leads to $\theta_{XM}=\pi$ \cite{a33,f1}. We obtain from numerical calculations that $\theta_{YM}=0$ and $\theta_{XM}=\pi$ for the first three bands (see Fig.~2a and Supplemental Materials for details).

The edge states at opposite boundaries indeed intersect at $K_x=\frac{\pi}{a_x}$, as verified by finite-element simulation (FES) (see Fig.~2c), which is consistent with the nontrivial Zak phase along the XM line, $\theta_{XM}=\pi$ \cite{j1,f1}. The edge states dispersion measured in experiments using the Fourier-transformed field scan method (see methods) agrees well with the dispersion from the FES.
The nonreciprocal photon flow along the edge channel, characterized by the difference between the forward and backward transmission, is shown in Fig.~2d. Visible nonreciprocal photon flow exists in the frequency window of 12-12.8~GHz, slightly broader than the topological PBG. A higher frequency window of nonreciprocal transmission is also visible, which is due to the chiral edge states in the higher PBG. The sign-change picture in Fig.~1b is confirmed by the FES of the edge states for the dislocation structure with a chunk of topologically trivial PC (see Fig.~2e): on the left (right) side of the dislocation the two edge waves are of opposite (the same) phases. Such a sign-change feature, being a smoking-gun signature of the Dirac mass domain-wall, is confirmed in our experiments as well (see Fig.~2f).

The experimental set-up for the dislocation structure and the measurement scheme is shown in Fig.~3a. The EM wave is excited through the feed-probe  near the bottom cladding and detected by the detect-probe  near the top cladding. The feed-probe  is fixed at a position near the dislocation, while the detection position is changeable. Both the amplitude and the phase of the local EM fields are scanned using a 2D mapping system in a frequency-resolved manner (see Methods).

Our FES study indicates that there is no localized state in the topologically trivial PBG between the second and the third bands (denoted as ``PBG I", 8.12-10.17~GHz), whereas there is one cavity mode in PBG II (topologically nontrivial PBG, 12.05-12.60~GHz) which is localized on the dislocation (see Fig.~3b). This observation confirms that the emergence of the mid-gap cavity mode is solely due to the dual-topology mechanism: without the dual-topology, the dislocation alone cannot induce 0D light-trapping. The EM field-profile of the cavity mode indicates strong light-trapping on the dislocation with a localization length $l_{loc}=1.0\lambda$ (estimated from $l_{loc}=\sqrt{{\cal A}/\pi}$ where ${\cal A}$ is the modal-area). Experimentally, we measure the transmission between two points which are located at different sides of the dislocation (inset of Fig.~3c). The transmission has only one peak in the PBG II (the shaded regions in Figs.~3c-3f), indicating only one localized mid-gap mode. The mode profile measured at the peak frequency in experiments is comparable with the field profile of the topological cavity mode from the FES (see the insets of Fig.~3c).

We now study the transmission and the field-profiles when the dislocation structure is strongly perturbed. For instance, we replace one of the YIG pillar close to the dislocation by a metallic pillar of the same size. As shown for the cases in Figs.~3d-3f, the topological cavity mode is found to be resilient against such perturbations. For all these cases, there is only one peak in the transmission spectrum in the PBG II, which reflects the robustness of the cavity mode and the topological light-trapping mechanism. One possible application in photonic circuits, a waveguide-coupler which couples two helical edge channels through the topological cavity mode, is demonstrated in the Supplemental Materials. 

Achieving topological light-trapping at $d-2$ dimensions in a $d$-dimensional photonic system opens up the possibility of realizing many new physical effects and applications, as well as answering several important open questions: whether dual-topology can be exploited to induce wave localization in nonlinear or non-Hermitian systems where lasing \cite{laser1,laser2,laser3}, parity-time symmetry and other important effects can emerge \cite{yd1}; what are the consequences when such topological cavity modes are coupled with quantum dots or other single-photon resources; can the dual-topology mechanism be extended to photonic quantum spin Hall insulators \cite{d1,d2,d3,d4,d5,d6,d7,d8} or valley Hall insulators \cite{valley} which can be realized by using all-dielectric structures and useful for optical frequency applications; how to extend to higher dimensions, such as a topological waveguide induced by a 1D dislocation line in a 3D photonic crystal. All these questions can lead to interesting physics and applications in the future.

\section*{{\bf Methods}}

\noindent {\bf Materials and sample fabrication.} \\
All the samples in this paper are fabricated by the low loss commercial YIG ferrite pillars. The saturation magnetization is measured as $4\pi M_s=1884$~Gauss by vibrating sample magnetometer and relative permittivity is retrieved as $15-0.003i$ by transmission/reflection method which can be treated as a constant at the microwave frequencies of interest (i.e., the PBG II). The fired ferrite is machined into pillars with a radius of $R=2$~mm and height $h=10$~mm. The topologically trivial PC is realized by reducing $a_x$ and $a_y$ to half of their original values, $d\to d_1=5$~mm, while keeping the same radius and height of the pillars.\\

\noindent {\bf Experimental setup and measurement.}\\
The measurement setup for the topological dislocation, as schematically shown in Fig.~3a, consists of vector network analyzer Agilent E8363A, a 2D mapping system, in a structure with top and bottom metallic cladding using aluminum plates. The upper (fixed) metallic plate has an area of $1\times 1$~m$^2$. The lower metallic plate (movable) has an area of $0.5\times 0.5$~m$^2$. The mapping area can be as large as $0.5\times 0.5$~m$^2$ when a single detect-probe is used. An array consisting of 364 permanent magnet  NdFeB pillars are embedded into a 3~mm thick aluminum plate (the lower metallic plate) in the sample with the dislocation. This plate works simultaneously as the metallic cladding as well as the external magnetic field source. Each NdFeB pillar is of radius 1.5~mm and height 3~mm. It can induce maximally 2800~Oe surface magnetic field. These NdFeB pillars apply one-to-one external magnetic field to the YIG pillars set between the magnet plate and top aluminum plate. Since the NdFeB pillars are outside of the metallic cladding, they do not affect the EM waves inside the cladding, except providing the magnetic fields. On average, the NdFeB pillars provide an external magnetic field of  about 900~Oe. The whole photonic structure is surrounded by microwave absorbers. The field-profiles are measured by scanning the EM fields through changing the position of the detect-probe. \\

\noindent {\bf Band structure and simulation.}\\ 
The band structure and all the simulations were calculated by using the commercial software COMSOL MULTIPHYSICS with the RF module.

\vskip 0.5cm

\noindent {\bf Acknowledgements}\\
FFL, QL, PC, RXW \& YP thank the support of National Natural Science Foundation of China (NSFC Grant Nos. 61671232, 61771237). HXW, ZX \& JHJ acknowledge supports from the National Natural Science Foundation of China (NSFC Grant No. 11675116) and the Soochow University. SJ acknowledges support from Natural Sciences and Engineering Research Council of Canada.

\noindent {\bf Author Contributions}\\
JHJ initiated the project. JHJ, YP, RXW \& SJ guided the research. HXW \& ZX designed the photonic architecture, performed theoretical analysis and calculations. FFL, QL, PC, RXW \& YP designed and conducted the experiments. All authors contributed to the analysis of results and the underlying mechanisms. JHJ, YP \& SJ wrote the manuscript.

\noindent {\bf Additional Information}\\
Supplementary information and movie are available in the online version of the paper. Reprints and permissions information are available online at www.nature.com/reprints. Correspondence and requests for data and materials should be addressed to JHJ (joejhjiang@hotmail.com) \& YP (ypoo@nju.edu.cn).

\noindent {\bf Competing Interests}\\
The authors declare no competing financial interests.

\vskip 2cm

{\bf Figures}\\

\begin{widetext}

\begin{figure}
\begin{center}
\includegraphics[width=13cm]{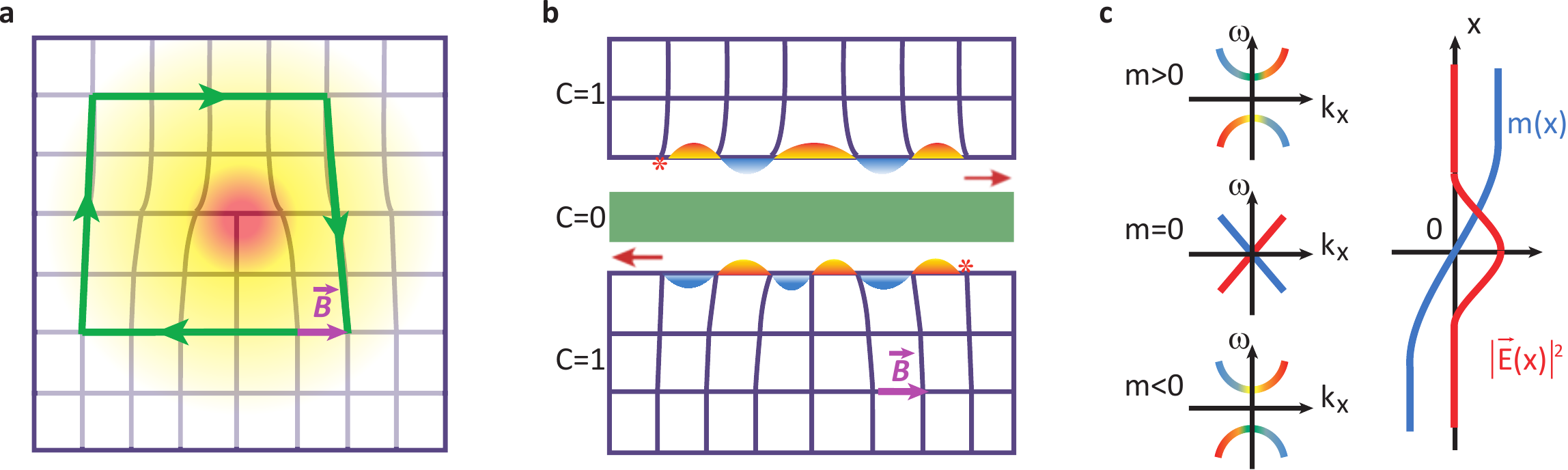}
\end{center}
\end{figure}

\begin{figure}
\begin{center}
\includegraphics[width=15cm]{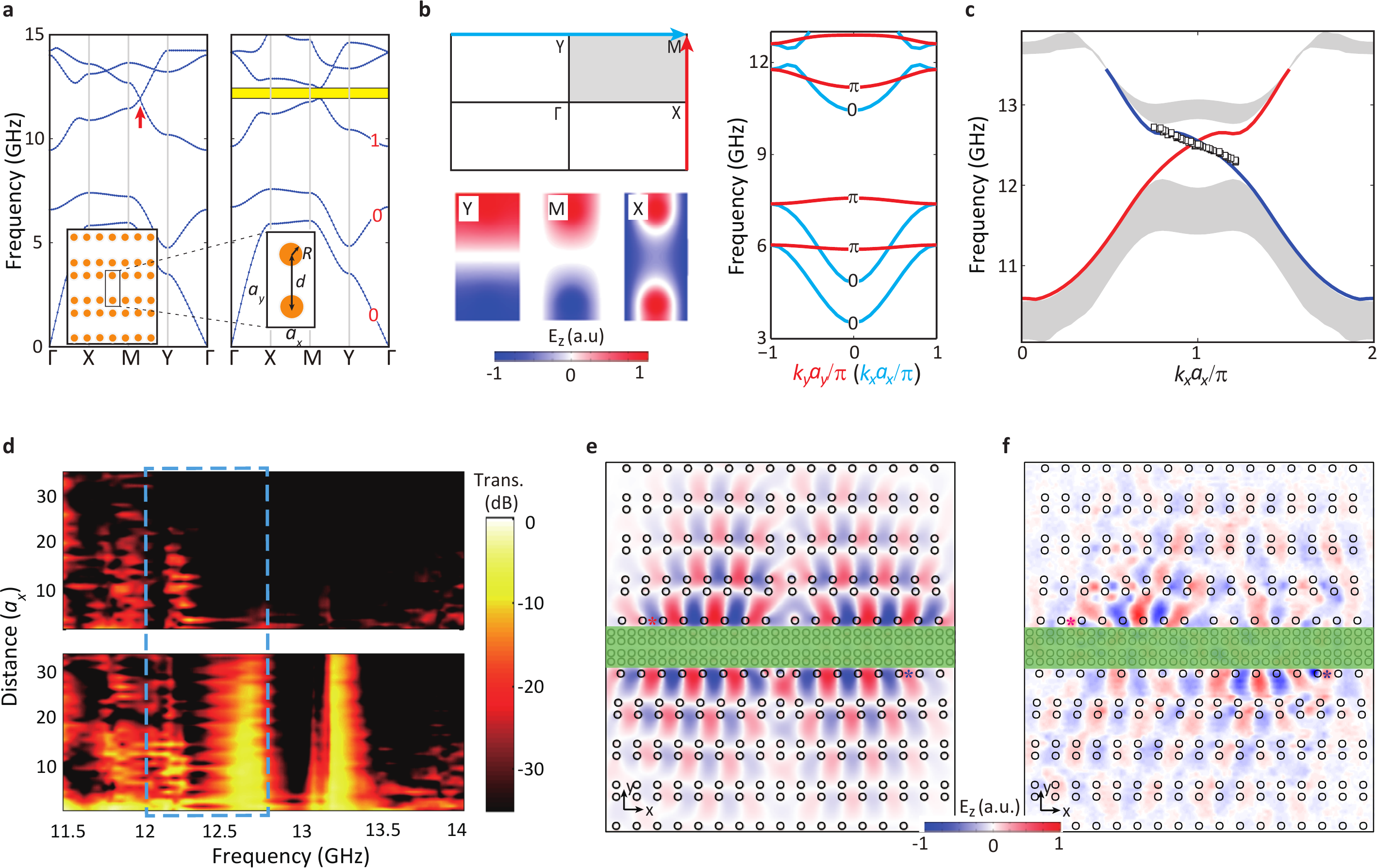}
\end{center}
\end{figure}

\begin{figure}
\begin{center}
\includegraphics[width=15cm]{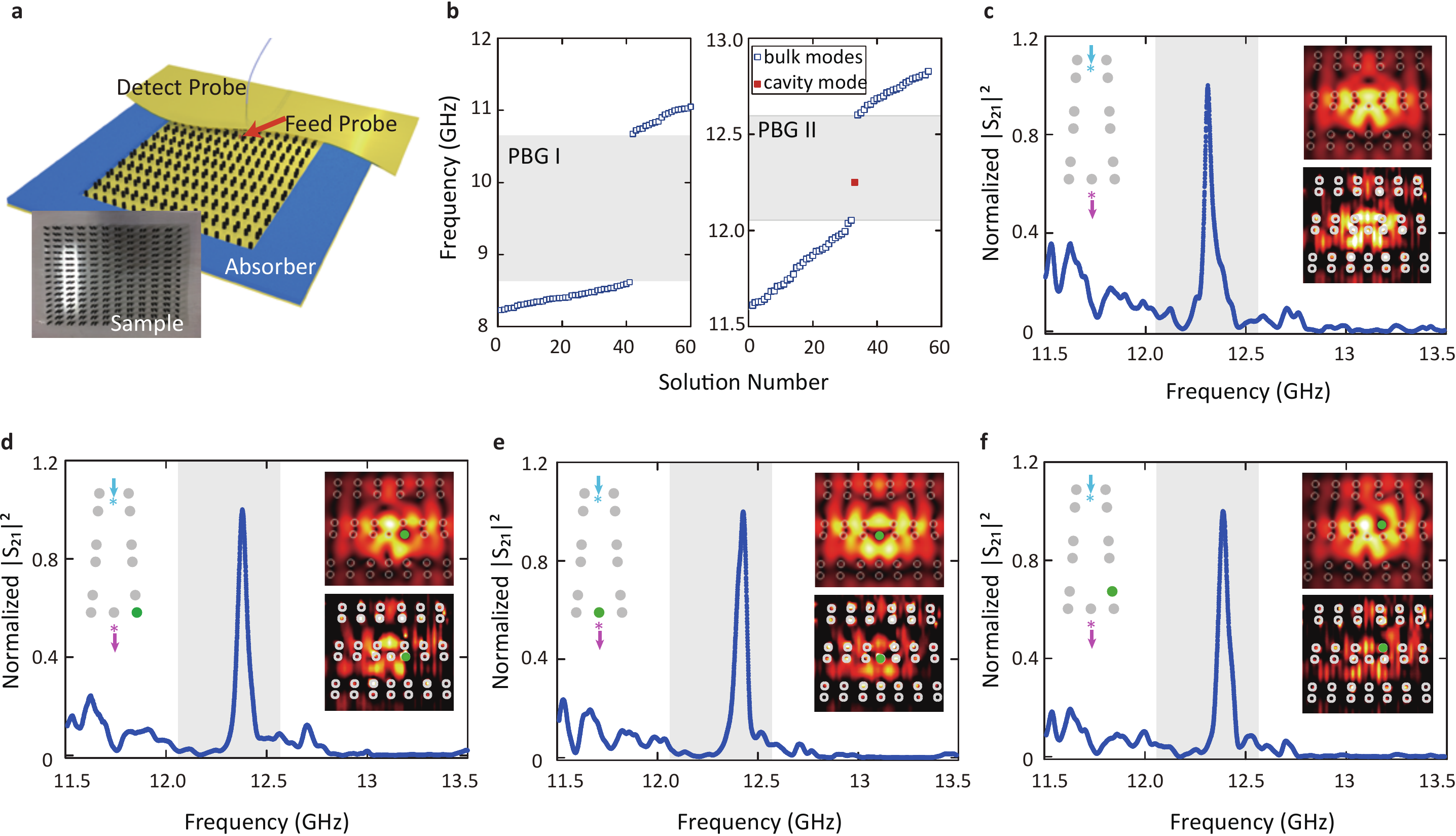}
\end{center}
\end{figure}

\end{widetext}

{\bf Figure Captions}:\\

{\bf Figure 1 | Light-trapping on a dislocation due to dual-topology.} {\bf a}, Schematic of a dislocation and the Burgers vector in a rectangular lattice. A photonic bound state is localized at the center of the dislocation. {\bf b}, Illustration of the ``cut-and-glue" technique. Inserting a chunk of PC with trivial topology (the green strip) into a topological PC with a dislocation yields two edge channels at opposite boundaries. Because of the additional lattice period in the lower edge, on the left (right) side of the dislocation there is a $\pi$ (no) phase difference between the two edge states. {\bf c}, The phase $\pi$ shift in the edge channels yields sign change in the Dirac mass across the dislocation, forming a Dirac mass domain-wall which results in a photonic Jackiw-Rebbi soliton mode.\\
 
{\bf Figure 2 | Topological photonic crystal: bulk, edge and dislocation.} {\bf a}, Design of the PC and PBG with nontrivial topology (Chern numbers labeled by red numbers). {\bf b}, Illustration of the BZ, the dispersion and the Zak phase along the XM (YM) line [the red (cyan) curves], and the field-profiles of the first band at X, Y and M points. {\bf c}, Edge states from the FES (red and blue curves) and experiments (square blocks, only at one edge). Bulk bands are represented by gray areas. {\bf d}, Nonreciprocal light propagation: the forward (upper panel) and backward (lower panel) transmission as functions of frequency and the distance between the source and the detection points along the edge. {\bf e} and {\bf f}, confirmation of the $\pi$ phase shift in the dislocation structure with a chunk of trivial PC in the FES ({\bf e}) and experiments ({\bf f}).\\

{\bf Figure 3 | Robust topological cavity mode on a dislocation.} {\bf a}, Experimental set-up: the PC with a dislocation is cladded above and below by aluminum plates and mounted onto a 2D mapping system. {\bf b}, Bulk and cavity modes from FES for PBG I (left) and PBG II (right). There is only one cavity mode in PBG II, whereas no cavity mode in PBG I. {\bf c}-{\bf f}, Transmission and field-profiles (upper insets: FES, lower insets: experiments). The left insets show the structure near the dislocation: the gray points represent the YIG pillars, the green points denote metallic pillars. The upper arrows indicate EM wave injection at the cyan asterisk point, while the lower arrows indicate EM wave detection at the purple asterisk point in transmission measurements. The peak frequencies in {\bf c}-{\bf f} are 12.37, 12.43, 12.50 and 12.43~GHz, respectively.\\

\end{document}